# SA-UNet: Spatial Attention U-Net for Retinal Vessel Segmentation

Changlu Guo[1,*], Márton Szemenyei[1], Yugen Yi[2,*], Wenle Wang[2], Buer Chen[1], Changqi Fan[1]

[1]Budapest University of Technology and Economics, Budapest, Hungary
Email: clguo.ai@gmail.com
[2]Jiangxi Normal University, Nanchang, China
Email: yiyg510@jxnu.edu.cn

*Abstract*— The precise segmentation of retinal blood vessels is of great significance for early diagnosis of eye-related diseases such as diabetes and hypertension. In this work, we propose a lightweight network named Spatial Attention U-Net (SA-UNet) that does not require thousands of annotated training samples and can be utilized in a data augmentation manner to use the available annotated samples more efficiently. SA-UNet introduces a spatial attention module which infers the attention map along the spatial dimension, and multiplies the attention map by the input feature map for adaptive feature refinement. In addition, the proposed network employs structured dropout convolutional blocks instead of the original convolutional blocks of U-Net to prevent the network from overfitting. We evaluate SA-UNet based on two benchmark retinal datasets: the Vascular Extraction (DRIVE) dataset and the Child Heart and Health Study (CHASE_DB1) dataset. The results show that the proposed SA-UNet achieves state-of-the-art performance on both datasets. The implementation and the trained networks are available on [Github[1]](https://github.com/clguo/SA-UNet).

## I. Introduction

Many diseases can be easily diagnosed and tracked by observing the fundus vascular system, because these diseases (such as diabetes and hypertension) can cause morphological changes in the blood vessels of the retina. Systemic microvascular and small vessel diseases are common pathological changes caused by diabetes, especially the fundus retinal vascular disease is the most vulnerable. Diabetic retinopathy (DR) is caused by diabetes [1]. If swelling of the blood vessels in the retina of a diabetic patient is observed, special attention is required. Patients with long-term hypertension may observe blood vessel curvature due to increased arterial blood pressure or vascular stenosis, which is called hypertensive retinopathy (HR) [2]. Retinal vessel segmentation is a key step in the quantitative analysis of fundus images. By segmenting the retinal blood vessels, we can obtain the relevant morphological information of the retinal blood vessel tree (such as the curvature, length, and width of the blood vessels) [3]. Moreover, the vascular tree of retinal vessels has unique characteristics that can be applied to biometric recognition [4], [5] as well. Therefore, accurate segmentation of retinal blood vessels is of great significance.

However, retinal blood vessels have numerous small and fragile blood vessels, and the blood vessels are closely connected, so the retinal blood vessel tree structure is rather complex. In addition, the difference between the blood vessel area and the background is not obvious, and the fundus image is also susceptible to uneven lighting and noise. The above reasons cause retinal blood vessel segmentation to be a challenging task.

In the past few decades, a large number of retinal blood vessel segmentation methods have been proposed, mainly divided into manual and automatic segmentation methods. The former is time-consuming and labor-intensive and requires extremely high professional skills of practitioners. The latter can reduce the burden of manual segmentation, so the research on automatic segmentation algorithms is of great significance. With the advancement of deep learning in recent years, it has gradually become the mainstream technology of retinal segmentation.

In the field of medical image segmentation, U-Net [6] is a common and well-known backbone network. Basically, U-Net consists of a typical downsampling encoder and upsampling decoder structure and a "skip connection" between them. It combines local and global context information through the encoding and decoding process. Due to the excellent performance of U-Net, many recent methods for retinal blood vessel segmentation are based on U-Net. Wang et al. [7] reported the Dual Encoding U-Net (DEU-Net) that remarkably enhances network's capability of segmenting retinal vessels in an end-to-end and pixel-to-pixel way. Wu et al. [8] proposed Vessel-Net, which first time uses a strategy that combines the advantages of the initial method and the residual method to perform retinal vessel segmentation. Zhang et al. [9] proposed AG-Net, which designed an attention mechanism called "Attention Guide Filter" to better retain structural information. Although these U-Net variants perform well, they inevitably make the network more complex and less interpretable.

In order to address these problems, we introduce spatial attention in U-Net and propose a lightweight network model, which we named Spatial Attention U-Net (SA-UNet). As shown by SD-Unet [10], using DropBlock [11] can effectively prevent overfitting of the network, so even small sample data-sets, such as retinal fundus images can be well trained. In addition, batch normalization (BN) can improve the convergence speed of the network [12]. Therefore, SA-UNet first employs a variant of structured dropout convolutional block integrating DropBlock and batch normalization (BN) to replace the original U-Net convolutional block. More importantly, the difference between vascular and non-vascular features in the retinal fundus image is not obvious, especially the small and marginal vascular areas. With the introduction of a small number of additional parameters, spatial attention can enhance important features (such as vascular features) and suppress unimportant features, thereby improving the network's representation ability. We evaluate SA-UNet on two public retinal fundus image datasets: DRIVE and CHASE_DB1. We first evaluate the newly introduced part of the network through ablation experiments. The experimental results show that the structured dropout convolutional block and the spatial attention we introduced are effective, and compared with the original U-Net and AG-Net, the proposed SA-UNet is very lightweight. Finally, compared with other existing state-of-the-art methods for retinal vascular segmentation, our proposed SA-UNet achieves state-of-the-art performance.

---

\* Corresponding authors
[1] https://github.com/clguo/SA-UNet

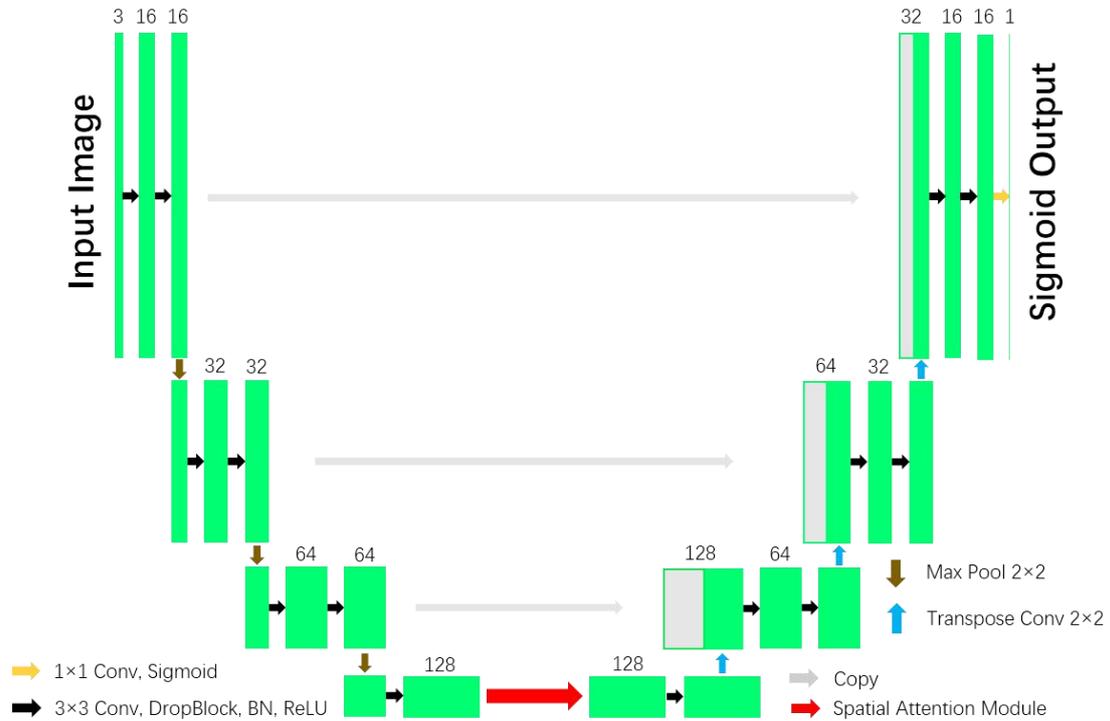

Fig. 1. Diagram of the proposed SA-UNet.

## II. METHODOLOGY

### A. Network Architecture

Fig. 1 shows the proposed SA-UNet with a U-shaped encoder (left side)-decoder (right side) structure. Every step of the encoder includes a structured dropout convolutional block and a 2×2 max pooling operation. The convolutional layer of each convolutional block is followed by a DropBlock, a batch normalization (BN) layer and a rectified linear unit (ReLU), and then the max pooling operation is utilized for down-sampling with a stride size of 2. In each down-sampling step, we double the number of feature channels. Each step in the decoder includes a 2×2 transposed convolution operation for up-sampling and halves the number of feature channels, a concatenates with the corresponding feature map from the encoder, which then followed by a structured dropout convolutional block. The spatial attention module is added between the encoder and the decoder. At the final layer, a 1×1 convolution and Sigmoid activation function is used to get the output segmentation map.

### B. Structured Dropout Convolutional Block

Although data augmentation is performed for the original datasets, serious overfitting is still observed during original U-Net training, as shown in Fig. 2 (left). Therefore a lightweight U-Net with 18 convolutional layers is employed as our basic architecture, but it still has over-fitting problem, as shown in Fig. 2 (middle). Motivated by the successful application of DropBlock in recent computer vision works [10], [11], [13], [19], we adopt DropBlock to regularize the network.

DropBlock, a structured form of dropout, can effectively prevent over-fitting problems in convolutional networks [10]. Its primary difference from dropout is that it discards contiguous areas from a feature map of a layer instead of dropping independent random units. Based on this, we construct a structured dropout convolutional block, that is, each convolutional layer is followed by a DropBlock, a layer of batch normalization (BN) and a ReLU activation unit, as shown in the right side of Fig. 3. Unlike the convolutional blo-ock of SD-Unet (as shown in the middle of Fig. 3), the structured dropout convolutional block introduces batch normalization (BN) to accelerate network convergence. We employ this structured dropout convolutional block instead of the original convolutional block of U-Net to build a U-shaped network as our "Backbone". Compared to the 23 convolutional layers of the original U-Net, our Backbone has only 18 convolutional layers, and as shown in Fig. 2. (left), the over-fitting problem is perfectly solved and accelerates the convergence of the network.

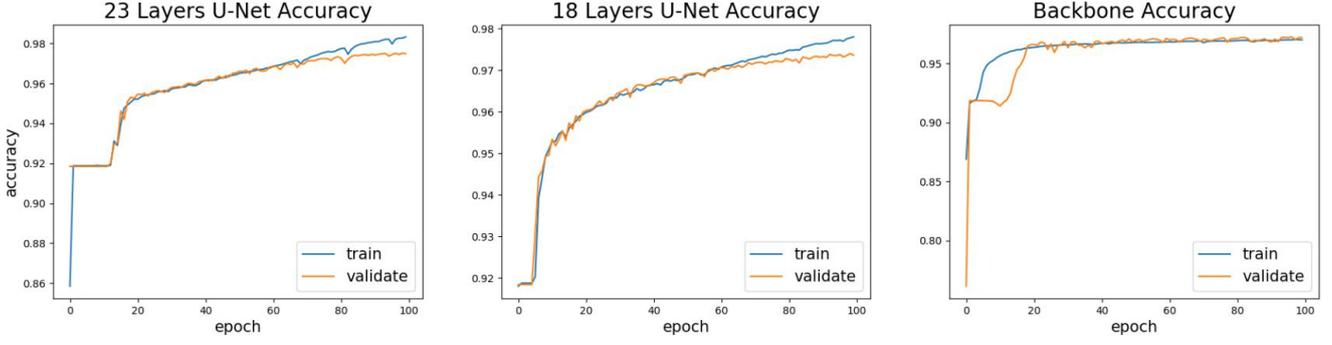

Fig. 2. Comparison of different models training 100 epochs on DRIVE.

### C. Spatial Attention Module (SAM)

The Spatial Attention Module (SAM) was introduced as a part of the *convolutional block attention module* for classification and detection [14]. SA uses the spatial relationship between features to produce a spatial attention map. To calculate spatial attention, SA first applies max-pooling and average-pooling operations along the channel axis and concatenate them to produce an efficient feature descriptor, as shown in Fig. 4. Formally, input feature $F \in R^{H \times W \times C}$ is fowarded through the channel-wise max-poling and average-pooling to generate outputs $F_{mp}^s \in R^{H \times W \times 1}$ and $F_{ap}^s \in R^{H \times W \times 1}$, respectively. Then a convolutional layer followed by the Sigmoid activation function on the concatenated feature descriptor is used to generate a spatial attention map $M^s(F) \in R^{H \times W \times 1}$. In short, the output feature $F^S \in R^{H \times W \times C}$ of spatial attention module is calculated as:

$$\begin{aligned} F^S &= F \cdot M^s(F) \\ &= F \cdot \sigma(f^{7 \times 7}([MaxPool(F); AvgPool(F)])) \\ &= F \cdot \sigma(f^{7 \times 7}([F_{mp}^s; F_{ap}^s])) \end{aligned} \quad (4)$$

Where $f^{7 \times 7}(\cdot)$ denotes a convolution operation with a kernel size of 7 and $\sigma(\cdot)$ represents the Sigmoid function.

## III. EXPERIMENTS

### A. Datasets

We evaluate our proposed SA-UNet on two public retinal fundus image datasets: DRIVE and CHASE DB1. The specific information on the two datasets is given in Table I. It should be noted that the original size of the two datasets is not suitable for our network, so we adjusted its size by zero padding around it, but the size is cropped to the initial size during evaluation. To augment the data, we adopt four data augmentation methods shown in the last column of Table I for both datasets, each of which generated three new images from an original image, that is, we augment the two original datasets from the original 20 training images to 256 images.

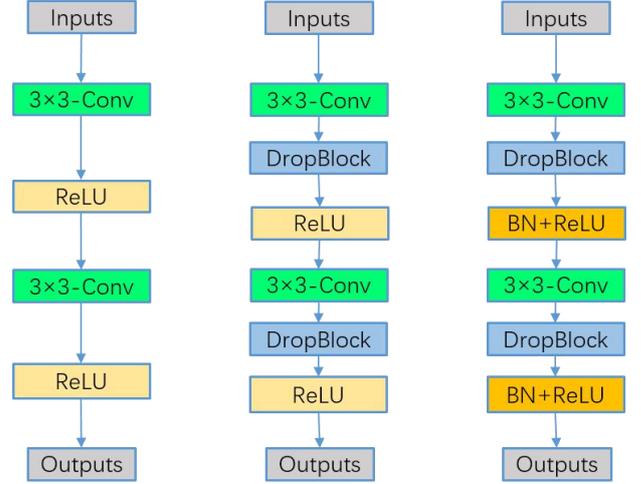

Fig. 3. Original U-Net block (left), SD-Unet block (middle), Structured dropout convolutional block (right)

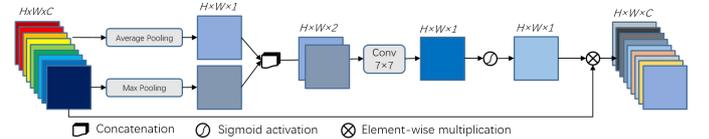

Fig. 4. Diagram of the Spatial Attention Module

### B. Evaluation Metrics

In order to evaluate our model, we compare the segmentation results with the corresponding ground truth and divide the results of each pixel comparison into true positive (*TP*), false positive (*FP*), false negative (*FN*), and true negative (*TN*). Then, the sensitivity (*SE*), specificity (*SP*), *F*1-score (*F1*), and accuracy (*ACC*) are used to evaluate the performance of the model. In retinal vessel segmentation, only 9%-14% of the pixels belong to the blood vessel, while other pixels are considered background pixels. The Matthews Correlation Coefficient (*MCC*) is suitable for performance measurement of binary classifications for two categories with different sizes. Therefore, the *MCC* value can help find the optimal setting for the vessel segmentation algorithm. *MCC* is defined as:

$$MCC = \frac{TP \times TN - FP \times FN}{\sqrt{(TP+FP) \times (TP+FN) \times (TN+FP) \times (TN+FN)}} \quad (5)$$

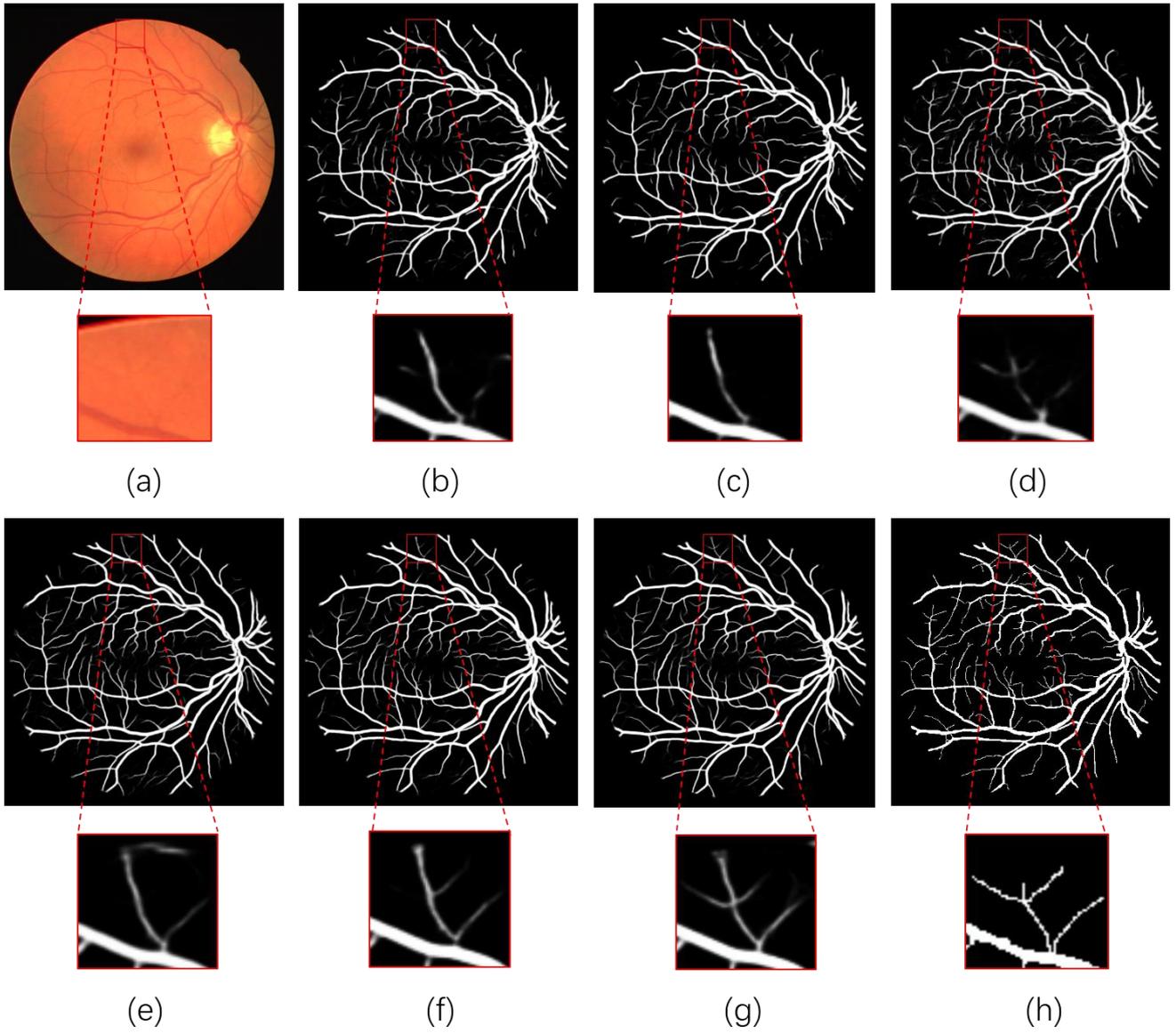

Fig. 5. (a) A test image from **DRIVE** dataset; (b) Segmentation result by **U-Net**; (c) Segmentation result by **U-Net+SA**; (d) Segmentation result by **AG-Net**; (e) Segmentation result by **SD-Unet**; (f) Segmentation result by **Backbone**; (g) Segmentation result by **SA-UNet**; (h) Corresponding ground truth segmentation.

The area under the ROC curve (AUC) can be used to measure the performance of the segmentation. If the AUC value is 1, it means perfect segmentation.

TABLE I.  THE SPECIFIC INFORMATION OF **DRIVE** AND **CHASE_DB1** DATASETS

| Datasets | **DRIVE** | **CHASE_DB1** |
|---|---|---|
| Obtained from | Dutch Diabetic Retinopathy Screening Program | Child Heart and Health Study |
| Total number | 40 | 28 |
| Train / Test number | 20 / 20 | 20 / 8 |
| Resolution (pixel) | 584×565 | 999×960 |
| Resize (pixel) | 592×592 | 1008×1008 |
| Augmentation methods | (1) Random rotation; (2) adding Gaussian noise; (3) color jittering; (4) horizontal, vertical and diagonal flips. | |

## C. Implementation Details

In order to monitor whether our network is overfitting, we randomly select 26 and 13 images in the DRIVE and CHASE DB1 augmented datasets as the validation set. As mentioned earlier, Fig. 2 shows the case of training 100 epochs on the DRIVE dataset. SA-UNet is trained from scratch using the augmented training set. For both datasets, the Adam optimizer and the binary cross entropy loss function are employed, and in order to keep the number of parameters small, the number of channels after the first convolutional layer is set to 16. The number of epochs is 150 and the learning rate of the first 100 epochs is 0.001, the last 50 epochs is 0.0001. The size of the discard blocks of DropBlock is set to 7.

Respectively, for DRIVE dataset, the batch size of the training is set to 8 and the dropout rate of DropBlock is set to 0.18. For CHASE DB1, the batch size is set to 4 and the dropout rates is 0.13.

The implementation is based on the public Keras with Tensorflow as the backend and all experiments are run on an NVIDIA TITAN XP GPU, which has 12 Gigabyte memory.

## IV. RESULTS

### A. Ablation Experiments

In order to prove that each component of the proposed SA-UNet can improve the performance of retinal vascular segmentation, ablation experiments were performed on DRIVE and CHASE_DB1 respectively. Tables II and III show the segmentation performance of U-Net, U-Net + SA, SD-Unet (i.e. U-Net + DropBlock), Backbone (i.e. SD-Unet + BN), and SA-UNet (i.e. Backbone + SA) from top to bottom, respectively. In addition, Table IV shows the parameter quantities of different models.

From the results, we could obtain several useful observations: (1) With only 98 parameters added, U-Net + SA has better performance compared with the U-Net, which proves the strategy of introducing spatial attention is effective. (2) In the case of using structured dropout convolutional block based on U-Net, the *ACC*, *AUC*, *F1* and *MCC* of the Backbone are 0.28% / 0.22%, 0.73% / 0.59%, 2.42% / 2.48%, and 2.48% / 2.64% higher than U-Net on DRIVE and CHASE_DB1 respectively, which demonstrates the effectiveness of adopting the newly constructed structured dropout convolutional block to build the Backbone. (3) Backbone has better performance compared to the original SD-Unet, although the number of parameters is increased slightly, which shows that adding the batch normalization (BN) can improve the network performance to a certain extent. (4) Finally, the proposed SA-UNet achieves the best performance on most metrics, and compared with AG-Net and the original U-Net with 23 convolutional layers, our SA-UNet has a much smaller amount of parameters, so for the task of retinal blood vessel segmentation, SA-UNet is a lightweight and effective network.

In Fig. 5, we show a test example on the DRIVE dataset, including the segmentation results obtained by U-Net, U-Net + SA, AG-Net, SD-Unet, Backbone and the proposed SA-UNet, and the corresponding ground truth. Compared with U-Net and U-Net + SA, AG-Net does have certain advantages in the segmentation of the edge structure, but at the intersection of small blood vessels, AG-Net is still not strong enough. SD-Unet ignores some edge and small vascular structures and there is even incorrect segmentation. The Backbone produces more accurate small vessel segmentation than the U-Net and SD-Unet, which proves the effectiveness of the Backbone constructed using structured dropout convolutional blocks. Compared with the Backbone, the SA-UNet proposed in this paper can produce more accurate segmentation results for small border blood vessels and retain more retinal blood vessel spatial structure, which proves that the spatial attention mechanism can highlight blood vessels and reduce the influence of background. In order to better observe the test results, we show more segmentation examples of U-Net, Backbone, and SA-UNet on DRIVE and CHASE_DB1 in Fig. 6 and Fig. 7, respectively.

TABLE II. ABLATION STUDIES ON **DRIVE** DATASET.

| Methods | SE | SP | ACC | AUC | F1 | MCC |
|---|---|---|---|---|---|---|
| U-Net | 0.7677 | 0.9857 | 0.9666 | 0.9789 | 0.8012 | 0.7839 |
| U-Net + SA | 0.7883 | 0.9845 | 0.9673 | 0.9809 | 0.8085 | 0.7909 |
| SD-Unet | 0.7978 | **0.9860** | 0.9695 | 0.9858 | 0.8208 | 0.8045 |
| Backbone | **0.8246** | 0.9832 | 0.9694 | 0.9862 | 0.8254 | 0.8087 |
| SA-UNet | 0.8212 | 0.9840 | **0.9698** | **0.9864** | **0.8263** | **0.8097** |

TABLE III. ABLATION STUDIES ON **CHASE_DB1** DATASET.

| Methods | SE | SP | ACC | AUC | F1 | MCC |
|---|---|---|---|---|---|---|
| U-Net | 0.7842 | 0.9861 | 0.9733 | 0.9838 | 0.7875 | 0.7733 |
| U-Net + SA | 0.7840 | **0.9865** | 0.9738 | 0.9852 | 0.7902 | 0.7763 |
| SD-Unet | 0.8297 | 0.9854 | **0.9756** | 0.9897 | 0.8109 | 0.7981 |
| Backbone | 0.8422 | 0.9844 | 0.9755 | 0.9897 | 0.8123 | 0.7997 |
| SA-UNet | **0.8573** | 0.9835 | 0.9755 | **0.9905** | **0.8153** | **0.8033** |

TABLE IV. AMOUNT OF PARAMETERS ON DIFFERENT MODELS.

| Models | Total | Trainable | Non-trainable |
|---|---|---|---|
| AG-Net | 9,335,340 | 9,335,340 | 0 |
| 23 Layers U-Net | 2,158,705 | 2,158,705 | 0 |
| 18 Layers U-Net | 535,793 | 535,793 | 0 |
| U-Net + SA | 535,891 | 535,891 | 0 |
| SD-Unet | 535,793 | 535,793 | 0 |
| Backbone | 538,609 | 537,201 | 1,408 |
| SA-UNet | 538,707 | 537,299 | 1,408 |

### B. Comparisons with state-of-the-art methods

Finally, we compare the performance of SA-UNet with other state-of-the-art methods currently applied in retinal vessel segmentation task. In Tables V and VI, we summarize the release year of different methods and the performance on DRIVE and CHASE_DB1 datasets. From the results, it can be concluded that SA-UNet has achieved the best performance on both DRIVE and CHASE_DB1. It achieves the highest sensitivity of 0.8212 / 0.8573, the highest accuracy of 0.9698 /

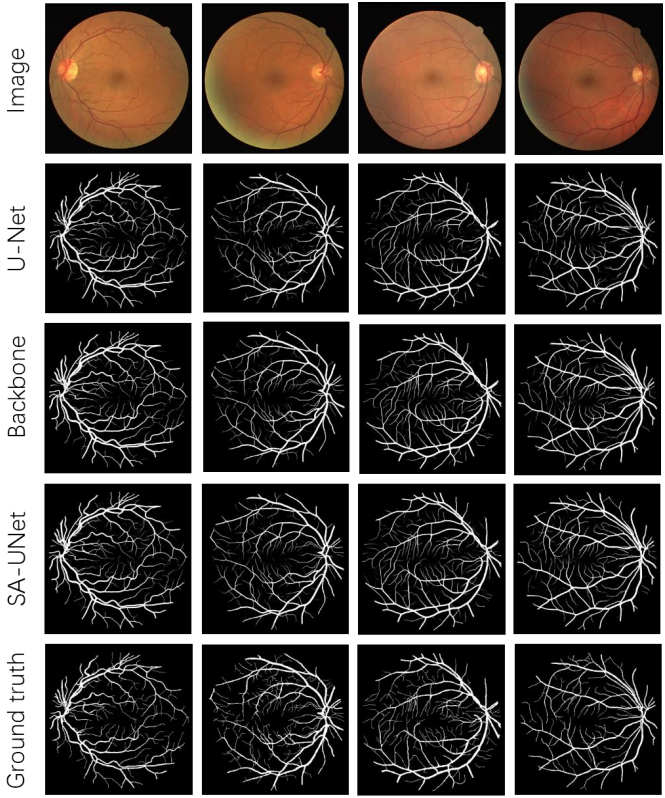

Fig. 6. Segmentation results on **DRIVE**

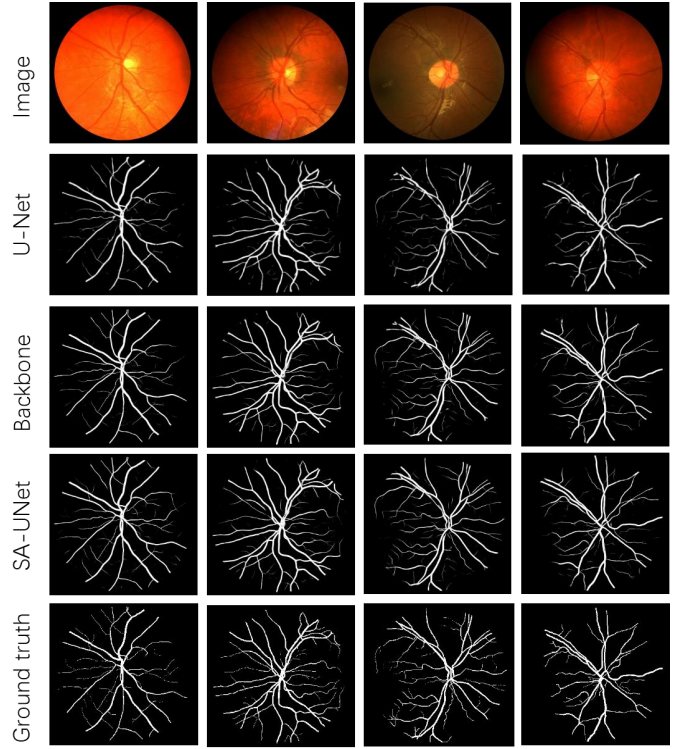

Fig. 7. Segmentation results on **CHASE_DB1**

0.9755, the highest AUC of 0.9864 / 0.9905, while the specificity is comparable with other methods. In addition, compared with the best performing AG-Net in the previous methods, SA-UNet has better segmentation performance at the intersection of small blood vessels, as shown in Fig. 5. Remarkablt, the parameter amount of SA-UNet is much smaller than that of AG-Net. The above results show that our proposed SA-UNet achieves state-of-the-art performance in the retinal vessel segmentation challenge.

TABLE V. RESULTS OF SA-UNET AND OTHER METHODS ON **DRIVE** DATASETS.

| Dataset | DRIVE | | | | |
|---|---|---|---|---|---|
| Metrics | Year | SE | SP | ACC | AUC |
| Liskowski et .al. [15] | 2016 | 0.7811 | 0.9807 | 0.9535 | 0.9790 |
| Orlando et. al. [16] | 2017 | 0.7897 | 0.9684 | 0.9454 | 0.9507 |
| Yan et. al. [17] | 2018 | 0.7653 | 0.9818 | 0.9542 | 0.9752 |
| MS-NFN [18] | 2018 | 0.7844 | 0.9819 | 0.9567 | 0.9807 |
| DEU-Net [7] | 2019 | 0.7940 | 0.9816 | 0.9567 | 0.9772 |
| Vessel-Net [8] | 2019 | 0.8038 | 0.9802 | 0.9578 | 0.9821 |
| AG-Net [9] | 2019 | 0.8100 | **0.9848** | 0.9692 | 0.9856 |
| **SA-UNet** | **2020** | **0.8212** | 0.9840 | **0.9698** | **0.9864** |

TABLE VI. RESULTS OF SA-UNET AND OTHER METHODS ON **CHASE_DB1** DATASETS.

| Datasets | CHASE_DB1 | | | | |
|---|---|---|---|---|---|
| Metrics | Year | SE | SP | ACC | AUC |
| Liskowski et .al. [15] | 2016 | 0.7816 | 0.9836 | 0.9628 | 0.9823 |
| Orlando et. al. [16] | 2017 | 0.7277 | 0.9712 | 0.9458 | 0.9524 |
| Yan et. al. [17] | 2018 | 0.7633 | 0.9809 | 0.9610 | 0.9781 |
| MS-NFN [18] | 2018 | 0.7538 | 0.9847 | 0.9637 | 0.9825 |
| DEU-Net [7] | 2019 | 0.8074 | 0.9821 | 0.9661 | 0.9812 |
| Vessel-Net [8] | 2019 | 0.8132 | 0.9814 | 0.9661 | 0.9860 |
| AG-Net [9] | 2019 | 0.8186 | **0.9848** | 0.9743 | 0.9863 |
| **SA-UNet** | **2020** | **0.8573** | 0.9835 | **0.9755** | **0.9905** |

## V. CONCLUSION

Most retinal fundus image datasets are typical small sample datasets, which can make training deep neural networks problematic. To enable learning, data augmentation is applied in an ambitious way, then a lightweight U-Net is used, but overfitting is still observed. Inspired by the successful application of DropBlock and batch normalization in convolutional neural networks, we replace the convolutional block of U-Net with a structured dropout convolutional block that integrates DropBlock and batch normalization as our Backbone. In addition, in the retinal fundus images, the difference between the blood vessel area and the background is not obvious, especially the edges and small blood vessels. To help the network learn these, we add a spatial attention module between the encoder and decoder of the Backbone and propose Spatial Attention U-Net (SA-UNet). The spatial attention can help the network focus on important features and suppress unnecessary ones to improve the network's representation capability. We evaluate SA-UNet on two publicly available retinal fundus image data including DRIVE and CHASE_DB1. The experimental results demonstrate that using structured dropout of convolutional blocks and the introducing spatial attention are effective, and by comparing with other state-of-the-art methods for retinal vessel segmentation, our lightweight SA-UNet achieves state-of-the-art performance. Because the vascular structure characteristics of the retinal image are similar, we conclude that SA-UNet is a general network and can be applied to other retinal vessel segmentation tasks.

ACKNOWLEDGMENT

This work is supported by the China Scholarship Council, the Stipendium Hungaricum Scholarship, and the National Natural Science Foundation of China under Grants 62062040.